\documentstyle[12pt]{article}

\begin{document}
\thispagestyle{empty}
\hspace{4.5in}{\vbox{\hbox{IFP-716-UNC}\hbox{July 1995}}}
\bigskip
\vspace{0.5in}
\begin{center}
{\LARGE\bf ~Treating Top Differently from Charm and Up*}\\

\vspace{7mm}
\large
Paul H. Frampton

\vspace{5mm}
{\it Institute of Field Physics, Department of Physics
and Astronomy,}\\
{\it University of North Carolina, Chapel Hill,
 NC, 27599--3255, USA}\\
\end{center}

\vspace{4.5in}

{\it *Contribution to the Workshop on Particle Theory and Phenomenology,
International Institute of Theoretical and Applied Physics (IITAP), Ames,
Iowa, May 22-26, 1995. }

\newpage

\centerline{\normalsize\bf TREATING TOP DIFFERENTLY FROM CHARM AND UP}
\baselineskip=16pt
\vspace*{0.3cm}

\centerline{\footnotesize PAUL H. FRAMPTON}
\baselineskip=13pt
\centerline{\footnotesize\it Institute of Field Physics, Department of Physics
and Astronomy,}
\baselineskip=12pt
\centerline{\footnotesize\it University of North Carolina, Chapel Hill, NC
27599-3255}
\centerline{\footnotesize E-mail: frampton@physics.unc.edu}
\vspace*{0.9cm}
\abstract{
It now appears phenomenologically that the third family may be essentially
different from the first two. Particularly the high value
of the top quark mass suggests a special role. In the standard model all three
families are
treated similarly [becoming exactly the same at asymptotically
high energies] so I need to
extend the model to accommodate the goal of
a really different third family.
In this article I describe not one but two such viable extensions, quite
different one
from another. The first is
the 331 model which predicts dileptonic gauge bosons.
In the second, using as a flavor symmetry a finite nonabelian dicyclic
$Q_{2N}$ group, I show how to derive quark
mass matrices with two arrangements of symmetric texture zeros which are
phenomenologically viable. Three other such
acceptable textures in the recent literature are unattainable in this
approach and hence disfavored. I assume massive vector-like
fermions and Higgs singlets transforming as
judiciously-chosen $Q_{2N}$ doublets and use the tree-level mass
generation mechanism of Froggatt and Nielsen.
}
\noindent

In this presentation, I shall address two extensions of the standard
model in which the third family is dealt with asymmetrically with respect to
the lightest two.
First, I shall describe the 331 Model and elaborate on its predictions
including the dilepton,
the neutrino masses and tests in hadronic and leptonic
colliders.
Second, after a review of finite nonabelian groups of order $\leq 31$ and the
question of their gauging and of their chiral anomalies, model building in that
direction is discussed
culminating in models
based on the dicyclic groups $Q_{2N}$.

\bigskip
{\bf 331 Model}
\bigskip

Family symmetry is usually taken to mean a horizontal symmetry, either global
or gauged,
under which the three families transform under some non-trivial representation.
The family symmetry is {\it broken} in order to avoid unobserved mass
degeneracies.
In this meaning of family symmetries, there is usually no explanation of {\it
why} there are
three families which are the input. Rather the hope is that the postulated
family
symmetry may explain the observed hierarchies:
\begin{equation}
m_u \ll m_c \ll m_t
\end{equation}
\begin{equation}
m_d \ll m_c \ll m_b
\end{equation}
\begin{equation}
\theta_{13} < \theta_{23} < \theta_{12}
\end{equation}

In the present model, the aim of family symmetry is indeed to attempt to
address such hierarchies {\it and} to explain why there are three
families. This may be a necessary first step to understanding hierarchies?

\bigskip

To introduce the 331 Model\cite{PHF} the following are motivating factors:

i) Consistency of a gauge theory (unitarity, renormalizability) requires
anomaly cancellation. This requirement almost alone is able to fix all
electric charges and other quantum numbers within one family
of the standard model. This accounts for charge quantization,
{\it e.g.} the neutrality of
the hydrogen atom, without the need for a GUT.

ii)This does not explain why $N_f > 1$ for the number of families but is
sufficiently impressive to suggest that $N_f = 3$ may be explicable by
anomaly cancellation in an extension of the standard model.
This requires that each
extended family have non-vanishing anomaly and that the three families
are not all treated
similarly.

iii) A striking feature of the mass spectrum in the SM is the top mass
suggesting that
the 3rd. family be treated differently and that the anomaly cancellation be
proportional
to: +1 +1 -2 = 0.

iv)There is a " -2 " lurking in the SM in the ratio of the quark electric
charges!

v)The electroweak gauge group extension from $SU(2)$ to $SU(3)$ will add five
gauge bosons. The adjoint of $SU(3)$ breaks into $8 = 3 + (2 + 2) + 1$ under
$SU(2)$.
The $1$ is a $Z^{'}$ and the two doublets are readily identifiable from the
leptonic triplet
or antitriplet $(e^{-}, \nu_e , e^{+})$ as {\it dilepton} gauge bosons
$( Y^{--} , Y^{-})$ with $L = 2$ and $( Y^{++} , Y^{+})$ with $L = -2$.
Such dileptons appeared first in stable-proton GUTs but there the fermions
were non-chiral and one needed to invoke mirror fermions; this is
precisely what is avoided in the 331 Model.  But it is true that the $SU(3)$
of the 331 Model has the same couplings to the {\it leptons} as that of the
leptonic
$SU(3)_l$
subgroup of $SU(15)$ which breaks to $SU(12)_q \times SU(3)_l$  .

\bigskip

Now I am ready to introduce the 331 Model in its technical details: the gauge
group
of the standard model is extended to $SU(3) \times SU(3) \times U(1) $ where
the electroweak
$SU(3)$ contains the standard $SU(2)$ and the weak hypercharge is a mixture of
$ \lambda_8 $
with the $U(1)$. The leptons are in the antitriplet $( e^-, \nu_e, e^+ )_L$ and
similarly
for the $\mu$ and $\tau$.

These antitriplets have $X = 0$ where $X$ is the new $U(1)$ charge. This can be
checked by
noting that the $X$ value is the electric charge of the central member of the
triplet or
antitriplet.

For the first family of quarks I use the triplet $( u, d, D )_L$ with $X =
-1/3$
and the right-handed counterparts in singlets. Similarly, the second family of
quarks is treated. For the third family of quarks, on the other hand, I use the
{\it antitriplet} $( T, t, b)_L$ with $X = +2/3$. The new exotic quarks D, S,
and T have
charges -4/3, -4/3 and +5/3 respectively.

It is instructive to see how this combination successfully cancels all chiral
anomalies:

The purely color anomaly $(3_L)^3$ cancels because QCD is vector-like.

The anomaly $(3_L)^3$ is non-trivial. Taking, for the moment, arbitrary
numbers $N_c$ of colors and $N_l$ of light neutrinos I find this anomaly
cancels only
if $N_c = N_l = 3$.

The remaining anomalies $(3_c)^2X$, $(3_L)^2X$, $X^3$ and $X(T_{\mu\nu}^2$
also cancel.

Each family separately has non-zero anomaly for $X^3$, $(3_L)^2X$ and
$(3_L)^3$;
in each case, the anomalies cancel proportionally to $+1 +1 -2$ between the
families.

\medskip

To break the symmetry I need several Higgs multiplets. A triplet $\Phi$ with $X
= +1$
and VEV $<\Phi>$ = $(0,0,U)$ breaks 331 to the standard 321 group, and gives
masses to
D, S, and T as well as to the gauge bosons Y and Z'.  The scale U sets the
range of the
new physics and I shall discuss more about its possible value.

The electroweak breaking requires two further triplets $\phi$ and $\phi'$ with
$X = 0$ and $X = -1$ respectively. Their VEVs give masses to d, s, t and to
u, c, b respectively. The first VEV also gives a contribution of an
antisymmetric-in-family type to the charged leptons. To complete a satisfactory
lepton mass matrix necessitates adding a sextet with $X = 0$.

\medskip

What can the scale $U$ be? It turns out that there is not only the lower
bound expected from the constraint of the precision electroweak data,
but also an upper bound coming from a group theoretical constraint within
the theory itself.

The lower bound on $U$ from $Z-Z'$ mixing can be derived from the
diagonalization of the mass matrix and leads to $M(Z') \geq 300GeV$.
The limit from FCNC (the Glashow - Weinberg rule is violated) gives a similar
bound; here the suppression is helped by ubiquitous $(1 - 4sin^2\theta)$
factors.

In these considerations, particularly with regard to FCNC, the special role
played by the third family is crucial; if either of the first two families
is the one treated asymmetrically the FCNC disagree with experiment.

\medskip

The upper bound on $U$ arises because the embedding of the standard $321$
group in $331$ requires that $sin^2\theta \leq 1/4$. When $sin^2\theta = 1/4$,
the $SU(2) \times U(1)$ group embeds entirely in $SU(3)$, and the coupling of
the $X$ charge in principle diverges. Because the phenomenological value is
close to 1/4 -
actually $sin^2\theta (M_Z) = 0.233$ - the scale $U$ must be less than about
$3TeV$ after scaling $sin^2\theta(\mu)$ by the renormalization group. Putting
some
reasonable upper bound on the $X$ coupling leads to an upper bound on the
dilepton
mass, for this 331 Model, of about $800GeV$ [ Here I have allowed one further
Higgs multiplet
- an octet].

\medskip

A very useful experiment for limiting the dilepton mass from below is
polarized muon decay. With the coupling parametrized as $V - \xi A$
where $\xi$ is a Michel parameter, the present limit on $\xi$ is $1 \geq \xi
\geq 0.997$
coming from about $10^8$ examples of the decay. This leads to a lower bound
$M(Y) \geq 300GeV$.

Since
\begin{equation}
(1 - \xi ) \sim (M_W/M_Y)^4
\end{equation}
I deduce that if $(1 - \xi )$ could be measured to an accuracy of $10^{-4}$
the limit would become $M_Y \geq 10 M_W$ and if to an accuracy $10^{-8}$
it would be $M_Y \geq 100M_W$. The first of these is within the realm
of feasability and certainly seems an important experiment to pursue.
The group at the Paul Scherrer Institute near Zurich (Gerber, Fetscher)
is one that is planning this experiment.

\bigskip

{\bf Neutrinos in the 331 Model}

\bigskip

In the minimal 331 Model as described so far, neutrinos are massless and
the model respects lepton number $\Delta L = 0$. Now I shall discuss
soft $L$ breaking for $M(\nu_i) \neq 0$.

Spontaneous breaking of $L$ would lead to a massless (triplet) majoron
in disagreement with experiment. Therefore I consider soft explicit breaking of
$L$.
The lepton families can be written $L_{i\alpha} = (l^-_i, \nu_i, l^+_i )_L$.
Among the Higgs scalars are the $H^{\alpha\beta}$ sextet and the
$\phi^{\alpha}$
triplet. The Yukawa couplings are:

\begin{equation}
h^{ij}_1L^i_{\alpha}L^j_{\beta}H^{\alpha\beta} +
h^{ij}_2L^i_{\alpha}L^j_{\beta}
{\overline{\phi}}_{\gamma}\epsilon^{\alpha\beta\gamma}  + h.c.
\end{equation}
The soft breaking of $L$ is in the triple Higgs couplings:
\begin{equation}
m_1H^{\alpha_1\beta_1}H^{\alpha_2\beta_2}H^{\alpha_3\beta_3}
\epsilon_{\alpha_1\beta_1\gamma_1}\epsilon_{\alpha_2\beta_2\gamma_2} +
m_2(H^{\alpha\beta} {\overline{\phi}}_{\alpha} {\overline{\phi}}_{\beta} +
h.c.)
\end{equation}

The neutrinos acquire mass from one-loop insertions of the soft breaking and
one
finds that provided the VEV $<H^{22}> = 0$ then there is the so-called cubic
see-saw formula:

\begin{equation}
M(\nu_i) = CM(l^-_i)^3/M_W^2
\end{equation}
where $l_i^-$ is the charged lepton corresponding to $\nu_i$ and C is a
constant calculable
in terms of various Yukawa couplings and Higgs masses
but whose absolute value is redundant in the sequel.  As an example, suppose I
adopt the
value for $\nu_{\tau}$ of $29.3eV$, an impressively precise value predicted by
Sciama's cosmology - obviously this is only an example! - then the other
neutrinos
have values $6.2meV$ and $690peV$ (where m is milli- and p is pico-).

\medskip

The $L$ breaking will also contribute to neutrinoless double beta decay but
the rate is around a billion ($10^9$) times below present experimental limits.

\medskip

The cubic see-saw with the cube of the charged lepton mass is numerically
quite similar to the more familiar quadratic see-saw with the up quark mass,
but since our present derivation does not involve a right-handed neutrino its
origin is conceptually quite independent. In any case, I can
fit the Hot Dark Matter and MSW requirements but not that for the atmospheric
neutrinos simultaneously just as for the Gell-Mann et al and Yanagida case.

\bigskip

{\bf Phenomenology of the 331 Model}

\bigskip

The dilepton can be produced in a hadron collider such as a $pp$ or $p
\overline{p}$
machine, or in a lepton collider such as $e^+e^-$ or $e^-e^-$.

For the hadron collider the $Y$ may be either pair produced or produced in
association
with an exotic quark [the latter carries $L = \pm 2$]. It turns out that the
associated
production is about one order of magnitude larger. These cross-sections are
calculated in the literature - for a pp collider of the type envisioned there
would
be at least $10^4$ striking events per year.

Surely the most dramatic way to spot a dilepton, however, would be to run a
linear collider in
the $e^-e^-$ mode and find a direct-channel resonance. A narrow spike at
between $300GeV$
and $800GeV$ would have a width at most a few percent of its mass and its decay
to $\mu^-\mu^-$ has no standard model background.

\medskip

{\bf Key Points of 331 Model:}

\bigskip

(i) The family symmetry can attempt to explain the fermion hierarchy and why
there
are three families.

(ii)In the 331 Model, the neutrino mass is either zero or proportional to
the cube of the charged lepton mass, depending on whether or not
one softly breaks $L$.

(iii)The dilepton (300GeV - 800GeV) could produce a narrow resonance in the
$e^-e^-$ mode.

\bigskip

\bigskip

{\bf Finite Groups as Family Symmetries}

\bigskip

As a new topic, let me turn to consideration of generic models of the type
where the symmetry group is $SM \times G$ with $SM$ the standard group
and $G$ is a finite group under which the families tranform under some
non-trivial
representation. This has already been studied for the abelian groups $Z_N$
and for certain non-abelian cases $S_3$ and $S_4$.

Before focusing in on specific groups, let me step back and look at all finite
groups of order
$g \leq 31$. [It is normal to stop at $g = 2^n - 1$ because $g = 2^n$ is always
so rich in groups.]

There are altogether 93 inequivalent such groups: 48 are abelian and the
remaining
45 non-abelian. Groups with $g \geq 32$ might well also be interesting but
surely
lower $g$ is simpler.

\bigskip

{}From any good textbook on finite groups\cite{books} one may find a tabulation
of
the number of finite groups as a function of the order g, the number of
elements in the group.

Amongst finite groups, the non-abelian examples have the advantage
of non-singlet irreducible representations which can be used to inter-relate
families. Which such group to select is based on simplicity: the minimum
order and most economical use of representations\cite{guts}.

Let me first dispense with the abelian groups. These are all made up from
the basic unit $Z_p$, the order p group formed from the $p^{th}$ roots
of unity. It is important to note that the the product $Z_pZ_q$ is identical
to $Z_{pq}$ if and only if p and q have no common prime factor.

If I write the prime factorization of g as:
\begin{equation}
g = \prod_{i}p_i^{k_i}
\end{equation}
where the product is over primes, it follows that the number
$N_a(g)$ of inequivalent abelian groups of order g is given by:
\begin{equation}
N_a(g) = \prod_{k_i}P(k_i)
\end{equation}
where $P(x)$ is the number of unordered partitions of $x$.
For example, for order $g = 144 = 2^43^2$ the value would be
$N_a(144) = P(4)P(2) = 5\times2 = 10$. For $g\leq31$ it is simple
to evaluate $N_a(g)$ by inspection. $N_a(g) = 1$ unless g contains
a nontrivial power ($k_i\geq2$) of a prime. These exceptions are:
$N_a(g = 4,9,12,18,20,25,28) = 2; N_a(8,24,27) = 3$; and $N_a(16) = 5$.
This confirms that:
\begin{equation}
\sum_{g = 1}^{31}N_a(g) = 48
\end{equation}
I shall not consider these abelian cases further, because all their irreducible
representations
are one-dimensional.

\bigskip

Of the nonabelian finite groups, the best known are perhaps the
permutation groups $S_N$ (with $N \geq 3$) of order $N!$
The smallest non-abelian finite group is $S_3$ ($\equiv D_3$),
the symmetry of an equilateral triangle with respect to all
rotations in a three dimensional sense. This group initiates two
infinite series, the $S_N$ and the $D_N$. Both have elementary
geometrical significance since the symmetric permutation group
$S_N$ is the symmetry of the N-plex in N dimensions while the dihedral group
$D_N$ is the symmetry of the planar N-agon in 3 dimensions.
As a family symmetry, the $S_N$ series becomes uninteresting rapidly
as the order and the dimensions of the representions increase. Only $S_3$
and $S_4$ are of any interest as symmetries associated with the particle
spectrum\cite{Pak}, also the order (number of elements) of the $S_N$ groups
grow factorially ($N!$) with N. The order of the dihedral groups increase only
linearly ($2N$) with N and their irreducible representations are all one- and
two- dimensional. This is reminiscent of the representations of the
electroweak $SU(2)_L$ used in Nature.

Each $D_N$ is a subgroup of $O(3)$ and has a counterpart double dihedral
group $Q_{2N}$, of order $4N$, which is a subgroup of the double covering
$SU(2)$ of $O(3)$.

With only the use of $D_N$, $Q_{2N}$, $S_N$ and the tetrahedral group T ( of
order
12, the even permutations subgroup of $S_4$ ) I find 32 of the 45
nonabelian groups up to order 31, either as simple groups or as
products of simple nonabelian groups with abelian groups:
(Note that $D_6 \simeq Z_2 \times D_3, D_{10} \simeq Z_2 \times D_5$ and $
D_{14} \simeq Z_2 \times D_7$ )

$$\begin{tabular}{||c||c||}   \hline
g & \\    \hline
$6$  & $D_3 \equiv S_3$\\  \hline
$8$ & $ D_4 , Q = Q_4 $\\    \hline
$10$& $D_5$\\   \hline
$12$&  $D_6, Q_6, T$ \\ \hline
$14$& $D_7$\\  \hline
$16$& $D_8, Q_8, Z_2 \times D_4, Z_2 \times Q$\\  \hline
$18$& $D_9, Z_3 \times D_3$\\  \hline
$20$& $D_{10}, Q_{10}$ \\  \hline
$22$& $D_{11}$\\  \hline
$24$& $D_{12}, Q_{12}, Z_2 \times D_6, Z_2 \times Q_6, Z_2 \times T$,\\  \hline
 & $Z_3 \times D_4, D_3 \times Q, Z_4 \times D_3, S_4$\\  \hline
$26$& $D_{13}$\\  \hline
$28$& $D_{14}, Q_{14}$ \\  \hline
$30$& $D_{15}, D_5 \times Z_3, D_3 \times Z_5$\\  \hline
\end{tabular}$$
There remain thirteen others formed by twisted products of abelian factors.
Only certain such twistings are permissable, namely (completing all $g \leq 31$
)

$$\begin{tabular}{||c||c||}   \hline
g & \\    \hline
$16$  & $Z_2 \tilde{\times} Z_8$ (two, excluding $D_8$), $Z_4 \tilde{\times}
Z_4, Z_2 \tilde{\times}(Z_2 \times Z_4)$
(two)\\  \hline
$18$ & $Z_2 \tilde{\times} (Z_3 \times Z_3)$\\    \hline
$20$&  $Z_4 \tilde{\times} Z_7$ \\   \hline
$21$&  $Z_3 \tilde{\times} Z_7$ \\    \hline
$24$&  $Z_3 \tilde{\times} Q, Z_3 \tilde{\times} Z_8, Z_3 \tilde{\times} D_4$
\\  \hline
$27$&  $ Z_9 \tilde{\times} Z_3, Z_3 \tilde{\times} (Z_3 \times Z_3)$ \\
\hline
\end{tabular}$$

It can be shown that these thirteen exhaust the classification of {\it all}
inequivalent finite groups up to order thirty-one\cite{books}.

Of the 45 nonabelian groups, the dihedrals ($D_N$) and double dihedrals
($Q_{2N}$), of order 2N and 4N respectively,
form the simplest sequences. In particular, they fall into subgroups of
$O(3)$ and $SU(2)$ respectively,
the two simplest nonabelian continuous groups.

For the $D_N$ and $Q_{2N}$, the multiplication tables, as derivable from the
character tables,
are simple to express in general. $D_N$, for odd N, has two singlet
representations $1,1^{'}$ and $m = (N-1)/2$
doublets $2_{(j)}$ ($1 \leq j \leq m$). The multiplication rules are:

\begin{equation}
1^{'}\times 1^{'} = 1 ; ~~~1^{'}\times 2_{(j)} = 2_{(j)}
\end{equation}
\begin{equation}
2_{(i)}\times 2_{(j)} = \delta_{ij} (1 + 1^{'}) + 2_{(min[i+j,N-i-j])}
+ (1 - \delta_{ij}) 2_{(|i - j|)}
\end{equation}
\noindent

For even N, $D_N$ has four singlets $1, 1^{'},1^{''},1^{'''}$ and $(m - 1)$
doublets
$2_{(j)}$ ($ 1 \leq j \leq m - 1$)where $m = N/2$ with multiplication rules:

\begin{equation}
1^{'}\times 1^{'} = 1^{''} \times 1^{''} = 1^{'''} \times 1^{'''} = 1
\end{equation}
\begin{equation}
1^{'} \times 1^{''} = 1^{'''}; 1^{''} \times 1^{'''} = 1^{'}; 1^{'''} \times
1^{'} = 1^{''}
\end{equation}
\begin{equation}
1^{'}\times 2_{(j)} = 2_{(j)}
\end{equation}
\begin{equation}
1^{''}\times 2_{(j)} = 1^{'''} \times 2_{(j)} = 2_{(m-j)}
\end{equation}
\begin{equation}
2_{(j)} \times 2_{(k)} = 2_{|j-k|} + 2_{(min[j+k,N-j-k])}
\end{equation}

\noindent
(if $k \neq j, (m - j)$)

\begin{equation}
2_{(j)} \times 2_{(j)} = 2 _{(min[2j,N-2j])} + 1 + 1^{'}
\end{equation}

\noindent
(if $j \neq m/2$ )

\begin{equation}
2_{(j)} \times 2_{(m - j)} = 2_{|m - 2j|} + 1^{''} + 1^{'''}
\end{equation}

\noindent
(if $j \neq m/2 $)

\begin{equation}
2_{m/2} \times 2_{m/2} = 1 + 1^{'} + 1^{''} + 1^{'''}
\end{equation}

\noindent
This last is possible only if m is even and hence if N is divisible by {\it
four}.\\

For $Q_{2N}$, there are four singlets $1$,$1^{'}$,$1^{''}$,$1^{'''}$ and
$(N - 1)$ doublets $2_{(j)}$ ($ 1 \leq j \leq (N-1) $).

The singlets have the multiplication rules:

\begin{equation}
1 \times 1 = 1^{'} \times 1^{'} = 1
\end{equation}
\begin{equation}
1^{''} \times 1^{''} = 1^{'''} \times 1^{'''} = 1^{'}
\end{equation}
\begin{equation}
 1^{'} \times 1^{''} = 1^{'''} ; 1^{'''} \times 1^{'} = 1^{''}
\end{equation}

\noindent
for $N = (2k + 1)$ but are identical to those for $D_N$ when N = 2k.

The products involving the $2_{(j)}$ are identical to those given
for $D_N$ (N even) above.

This completes the multiplication rules for 19 of the 45 groups. When needed,
rules for the other groups will be derived.

Since I shall be emphasizing the groups $Q_{2n}$, let me be more explicit
concerning it and its destinction
from $D_{2n}$.

 $Q_{2n}$ is defined by the equations:
\begin{equation}
A^{2n} = E
\end{equation}
\begin{equation}
B^2 = A^n
\end{equation}
\begin{equation}
ABA = B
\end{equation}

I can find an explicit matrix representation in the form:

\begin{equation}
A = \left( \begin{array}{cc} cos\theta & sin\theta \\
-sin\theta & cos\theta \end{array} \right)
\end{equation}
with $\theta = \pi/n$. This gives then:
\begin{equation}
A^n = \left( \begin{array}{cc} cos (n\theta) & sin (n\theta) \\
-sin (n\theta) & cos (n\theta) \end{array} \right) \\
= \left( \begin{array}{cc} -1 & 0 \\
0 & -1 \end{array} \right)
\end{equation}
B is chosen as:
\begin{equation}
B = \left( \begin{array}{cc} i & 0 \\
0 & -i \end{array} \right)
\end{equation}

For $D_{2N}$, on the other hand the choice of B is replaced by:
\begin{equation}
B = \left( \begin{array}{cc} 1 & 0 \\
0 & -1 \end{array} \right)
\end{equation}
so that $B^2 = A^n = +1$ instead of $- 1$.

{}From these matrices one can deduce, for example, the geometrical
interpretation that whereas $D_6$ is the full
dihedral [{\it i.e.} two-sided] symmetry of a planar hexagon in $O(3)$, $Q_6$
is the full $SU(2)$
symmetry of an equilateral triangle when rotation by $2 \pi$ gives a sign $(-
1)$ and a rotation by
$4 \pi$ is the identity transformation.

\bigskip
\bigskip

{\bf Anomalies and Model Building.}

\bigskip

The models I shall consider have a symmetry comprised of the standard model
gauge group
$SU(3)_C \times SU(2)_L \times U(1)_Y $ producted with a nonabelian finite
group G.

If G is a global (ungauged) symmetry, there are problems if the spacetime
manifold is
topologically nontrivial since it has been shown that any such global
symmetry is broken in the presence of wormholes\cite{global}. From a Local
viewpoint (Local with a capital
means within a flat spacetime neighbourhood) the distinction between a global
and local (gauged) finite symmetry does not exist. The distinction exists only
in a
Global sense (Global meaning pertaining to topological aspects of the
manifold).
In a flat spacetime, gauging a finite group has no meaning. In the
presence of wormholes, expected from the fluctuations
occurring in quantum gravity, gauging G is essential. The mathematical
treatment of such a gauged finite group has a long history\cite{flat}.

In order to gauge the finite group G, the simplest procedure is to gauge
a continuous group H which contains G as a subgroup, and then to spontaneously
break H by choice of a Higgs potential. The symmetry breaking may occur at a
high
energy scale, and then the low energy effective theory will not contain any
gauge potentials or gauge bosons; this effective theory is, as explained above,
Locally identical
to a globally-invariant theory with symmetry G.

For example, consider G = $Q_6$ and H = $SU(2)$. I would
like to use only one irreducible representation $\Phi$ of $Q_6$
in the symmetry-breaking potential $V(\Phi)$. The irreps. of $Q_6$
are $1, 1', 1^{''}, 1^{'''}, 2, 2_S$. The $1^{''}, 1^{'''}$ and
$2_S$ are spinorial and appear in the decompositions only of
$2, 4, 6, 8 ....$ of $SU(2)$. Since $\Phi$ must contain the $1$
of $Q_6$ I must choose from the vectorial irreps. $3, 5, 7, 9 ...$
of $SU(2)$. The appropriate choice is the $7$ represented by a symmetric
traceless
third-rank tensor $\Phi_{ijk}$ with $\Phi_{ikk} = 0$.

For the vacuum expectation value, I choose
\begin{equation}
<\Phi_{111}> = +1; <\Phi_{122}> = -1
\end{equation}
and all other unrelated components vanishing. If I look for the $3\times3$
matrices $R_{ij}$ which
leave invariant this VEV I find from choices of indices in
\begin{equation}
R_{il}R_{jm}R_{kn}<\Phi_{lmn}> = <\Phi_{ijk}>
\label{eq18}
\end{equation}
that $R_{31} = R_{32} = 0$ (Use $<\Phi_{3ij}\Phi_{3ij}> = 0$) and that $R_{33}
= \pm1$. Then I find $(R_{11})^3 - 3R_{11}(R_{12})^2 = 1$ (Use $l = m = n = 1$
in $(\ref{eq18})$). This means that if $R_{11} = \cos\theta$ then $\cos3\theta
= 1$ or $\theta_n = 2\pi n/3$. So the elements of $Q_6$ are $A = R_3(\theta_1),
A^2, A^3$ and $B, BA, BA^2$ where $B =$ diag$(i, -i -i)$.

More generally, it can be shown that to obtain $Q_{2N}$ one must use an $N^{\rm
th}$ rank
tensor because one finds for the elements $R_{11}$ and $R_{12}$:
\begin{equation}
\sum_{p=0}^{[N/2]}(-1)^p{N\choose2p} (R_{11})^{N-2p}(R_{12})^{2p} =
\cos N\theta = 1
\end{equation}

If the group H is gauged, it must be free from anomalies. This entails several
conditions which must be met:

(a) The chiral fermions must fall into complete irreducible representations
not only of G but also of H.

(b) These representations must be free of all H anomalies including $(H)^3$,
$(H)^2Y$;
for the cases of H = $O(3), SU(2)$ only the latter anomaly is nontrivial.

(c) If H = $SU(2)$, there must be no global anomaly.

The above three conditions apply to nonabelian H. The case of an abelian H
avoids (a) and (c) but gives rise to additional mixed anomalies in (b).

For nonabelian H, conditions (b) and (c) are straightforward to write down and
solve.
Condition (a) needs more discussion. I shall focus on the special cases
of $O(3) \supset D_N$
and $SU(2) \supset Q_{2N}$.

For $O(3)$ the irreps. are ${\bf 1,3,5,7,....}$ dimensional. $D_N$ has irreps.
(for even $N = 2m$) $1, 1^{'}, 1^{''}, 1^{'''}$ and $2_{(j)} (1 \leq j \leq (m
- 1))$ and these correspond
to:
\begin{equation}
O(3):  {\bf 1}  \rightarrow 1 ; {\bf 3} \rightarrow (1^{'} + 2_{(1)})
\end{equation}
\noindent
and so on. The same situation occurs for odd N with irreps. $1, 1^{'}$ and
$2_{(j)} ( 1 \leq j \leq (N - 1)/2)$.

For $SU(2) \supset Q_{2N}$ the corresponding breakdown is:
\begin{equation}
{\bf 1} \rightarrow 1; {\bf 2} \rightarrow 2_{S(1)} ; {\bf 3} \rightarrow 1^{'}
+ 2_{(1)}
\end{equation}
\noindent
and so on, where the doublets of $Q_{2N}$, $2_{(1)}$ and $2_{S(1)}$,
are defined by Eq. (35).

These are the principal splittings of a continuous group irrep.
into finite subgroup irreps. I shall
need in my discussions of model building.

\bigskip

\bigskip

{\bf Applications of $Q_{2N}$}

\bigskip

There has recently been considerable interest in the structure
of the quark mass matrices, particularly in the idea of postulating
texture zeros in grand unified theories, with a view to
obtaining relations
between the masses and mixing angles\cite{GUTS1}, \cite{GUTS2}, \cite{GUTS3}.
 A list of  phenomenologically
viable
quark mass matrices bearing  a maximum number of symmetric
texture zeros
 was presented in [11]. The possibility of constructing such
mass matrices from a
scheme of gauged flavor symmetry has been
considered in [12] and [13].

There has also been considerable activity in the use of finite non-abelian
groups as flavor
symmetries\cite{finite1},\cite{finite2},\cite{finite3}, with a view to
generating the mass hierarchy.

Here I attempt a
synthesis of these two approaches and construct desirable quark mass
matrix
textures by using a nonabelian flavor symmetry (specifically a dicyclic
group $Q_{2N}$) together with the Froggatt-Nielsen mechanism of mass
generation\cite{FN}.

The particular choices are given in [21]. A generalization
(and simplification) to include supersymmetric grand unification
is given in [22,23].

While U(1) flavor symmetry constructions for quark mass matrices
with {\it nonsymmetric} hierarchical textures have  been attempted
\cite{nonsym},  the full list of such phenomenologically viable quark mass
matrices
is not yet available. My approach does not {\it a priori} give a symmetric
texture. However, as a first attempt, I consider here only the possibility
of constructing the symmetric texture patterns presented in [11]. The general
case of non-symmetric textures would  naturally be a very interesting next
step.

The use of $Q_{2N}$ as a finite flavor group has been discussed in detail in
[16]. I recall here that the irreducible representations of $Q_{2N}$ are four
singlets $1, 1^{'}, 1^{''}, 1^{'''}$ and $(N - 1)$ doublets $2_{(k)}$,
with $1 \leq k \leq (N - 1)$. Most important for my purposes are the
products:
\begin{equation}
2_{(k)} \times 2_{(l)} = 2_{(|k-l|)} + 2_{(min\{k+l,2N-k-l\})}
\end{equation}
where, in a generalized notation, $2_{(0)} \equiv 1 + 1^{'}$ and $2_{(N)}
\equiv 1^{''} + 1^{'''}$.

I assign the quarks to $Q_{2N}$ representations as follows:
$$\begin{array}{cccc}
\left. \begin{array}{c} \left( \begin{array}{c} t \\ b \end{array} \right)_{L}
\\
\left( \begin{array}{c} c \\ s \end{array} \right)_{L} \end{array}  \right\} &
2_{(2)}&
 \begin{array}{c} \left. \begin{array}{c} t_R\\c_R \end{array} \right\} \\
\left. \begin{array}{c} b_R\\s_R \end{array} \right\} \end{array}&
\begin{array}{c} 2_{(2)}\\
2_{(1)} \end{array} \\
\left( \begin{array}{c} u \\ d \end{array} \right)_{L}  &
1^{'}&\begin{array}{c} u_R\\d_R
\end{array} & \begin{array}{c} 1^{'}\\1 \end{array} \\
\end{array}$$

When I embed the finite spinorial $Q_{2N}$ into its continuous progenitor
$SU(2)$,
$1, 2_{(1)}$ and $(1^{'} +2_{(2)})$ correspond respectively to the singlet,
doublet and triplet representations. The above quark assignment is thus
anomaly-free if the leptons are assigned to: $$\begin{array}{cccc}
\left. \begin{array}{c} \left( \begin{array}{c} \nu_{\tau} \\ \tau^{-}
\end{array} \right)_{L} \\
\left( \begin{array}{c} \nu_{\mu} \\ \mu^{-} \end{array} \right)_{L}
\end{array}  \right\} & 2_{(1)}&
\left.  \begin{array}{c}  \tau^{+}_L\\ \mu^{+}_L \end{array} \right\}
&  2_{(2)} \\
\left( \begin{array}{c} \nu_{e} \\ e^{-} \end{array} \right)_{L}  & 1& e^{+}_L
 & 1^{'} \\
\end{array}$$

I shall not consider lepton masses further here. For the mass textures of the
quarks I postulate heavy
vector-like fermions and singlet Higgs and assume the quark masses
arise from tree graphs as in [17].

As the first of two successful examples, I demonstrate how to
derive the five-zero texture in Eqs. (37) and (38) below. [Note that no texture
with the
maximum number of six texture zeros can be phenomenologically viable.]

\begin{equation}
M_{u} = \left(\begin{array}{ccc}
0 & 0 & \lambda^{4} \\
0 & \lambda^{4} & \lambda^{2} \\
\lambda^{4} & \lambda^{2} & 1
\end{array}\right)
\end{equation}
\begin{equation}
M_{d} = \left(\begin{array}{ccc}
0 & \lambda^{4} & 0 \\
\lambda^{4} & \lambda^{3} & 0 \\
0 & 0 & 1
\end{array}\right)
\end{equation}

For these matrices I have suppressed all coefficients of order one since at the
present
stage I am satisfied to derive only the correct orders in
$\lambda$ for each entry.

The standard Higgs scalar doublets of $SU(2)_L$ are taken as a $2_{(4)}$ of
$Q_{2N}$ coupling
to the up quarks, and a $2_{(3)}$ coupling to the down quarks. I assume
these get VEVS that break $Q_{2N}$ and give mass only to the third family. For
the up quark mass matrix the entry $(M_u)_{33}$ is
of order 1 from the coupling $t_L(2_{(2)})t_R(2_{(2)})H_u(2_{(4)})$. At
leading order, all other entries $(M_u)_{ij}$ vanish. Similarly,
$b_L(2_{(2)})b_R(2_{(1)})H_d(2_{(3)})$ gives $(M_d)_{33}$ of order 1 and no
other $(M_d)_{ij}$.

To obtain the other entries in Eqs. (37) and (38) at order $\lambda ^n
(\lambda \sim sin\theta_{C} \sim 0.22$
where $\theta_{C}$ is the Cabibbo angle),  I
introduce a list of vector-like quark doublets
$ Q_i(2_{(i)}) (i=6, 7, 10, 13, 14),$
singlets
$U_i(2_{(i)}) (i=6, 10, 14)$
and
$D_i(2_{(i)}) (i=4, 17),$
 bearing the same standard model quantum numbers as
 $Q_L, u_R$ and $d_R$ respectively, together with standard model
singlet Higgses
$S_i(2_{(i)}),(i= 5, 8, 13, 14, 17, 20)$. Although this set of $Q_{2N}$
doublets seems long and {\it ad hoc}, it is highly constrained (see below).
Since I have assumed heavy particles in doublets up to $2_{20}$ the
flavor group, of order 84, is $Q_{42}$.

I choose a set of bases and label the two states in the heavy fermion $Q_{2N}$
doublets as $2_{(i)+}$ and $2_{(i)-}$, which lie
respectively in the third and second family direction.  The
$H_u$ VEV then allows only the six couplings:
\[	t_L<H_u>U_{6+}; \; Q_{6+}<H_u>t_R; \; Q_{6+}<H_u>U_{10+}; \]
\[  Q_{10+}<H_u>U_{6+};\; Q_{10+}<H_u>U_{14+}; \; Q_{14+}<H_u>U_{10+}.  \]
and the $H_d$ VEV only the two couplings:
\[	Q_{7+}<H_d>D_{4+};\; Q_{14+}<H_d>D_{17+}.\]
The $S_i$ VEVS may then be chosen to give certain vertices such as :
$U_{6+}^{\dag}<S_{8+-}>c_R,$ \\   $U_{10+}^{\dag}<S_{8+-}>c_R$, and others.
I define:
\begin{equation}
\lambda^2 = \frac{<S_i>}{M_{even}}
\end{equation}
\begin{equation}
\lambda = \frac{<S_i>}{M_{odd}}
\end{equation}
where $M_{even}$ and $M_{odd}$ denote
the mass of a heavy fermion in $Q_{2N}$ representation $2_{(k)}$ for being
$k$ even and odd respectively. Note that Eqs. (39) and (40) are acceptable
because
the k-even and
k-odd doublets occur independently in the irreducible representations of
the covering $SU(2)$ in the sense that the k-even doublets appear only in
vectors of SU(2) and the k-odd doublets appear only in spinors.

I have now all the ingredients of the model. In the low energy efffective
field theory, after integrating out the heavy fermions\cite{EFT}, I have
the tree level quark mass matrices having the structure of the
model denoted by roman numeral V in ref.\ \cite{RRR},
namely those exhibited in Eqs.\~(37)
and (38) above.

For instance, the  $(M_u)_{32}$ entry is given by the Froggatt-Nielsen
tree graph (shown in Fig.(1a)) correponding to the operator couplings
\[  	t_L <H_u> ( U_{6+} U_{6+}^{\dag} ) <S_{8+-}> c_R
 = \lambda ^2 <H_u> t_L c_R;
\]
while $(M_u)_{13}$ is given by the graph of Fig. (1c) corresponding to:
\[ 	t_L <H_u> ( U_{6+} U_{6+}^{\dag} )
<S_{8+-}> (U_{14-}U_{14-}^{\dag})  <S_{14-}>u_R\]
\[ = \lambda ^4 <H_u>  t_L u_R; \]
and $(M_d)_{22}$ is given by Fig.\ (2a) corresponding to:
\[ 	s_L <S_{5+-}> (Q_{7+}^{\dag}Q_{7+}) <H_d> (D_{4+}
D_{4+}^{\dag})<S_{5+-}>s_R\]
\[ =  \lambda^3 <H_d> s_L s_R.  \]
The other entries in $M_u$ and $M_d$ are derived similarly; some further
examples
are shown in Fig. (1b) for $M_u$ and Figs. (2b) and (2c) for $M_d$.

%
%
%

In the construction  of the model, I followed a systematic procedure and were
surprised
to realize that it is highly non-trivial if any consistent model
can be constructed at all. The difficulty is not only to derive the correct
texture zeros but also to avoid unwanted entries at too low an order in
$\lambda$.
I find only two consistent models, the above model and one alternative
summarized below.

The mass matrices for the alternative model are:

\begin{equation}
M_{u} = \left(\begin{array}{ccc}
0 & \lambda^{6} & 0 \\
\lambda^{6} & \lambda^{4} & \lambda^{2} \\
0 & \lambda^{2} & 1
\end{array}\right)
\end{equation}

\begin{equation}
M_{d} = \left(\begin{array}{ccc}
0 & \lambda^{4} & 0 \\
\lambda^{4} & \lambda^{3} & 0 \\
0 & 0 & 1
\end{array}\right)
\end{equation}

Note that the $M_{d}$ matrix is the same as in my first example but
$M_{u}$ is changed; as before, I neglect coefficients of order unity. The
$Q_{2N}$
assignments for the quarks and leptons are the same as they were previously.

The entries in Eqs.\ (41) and (42) at order $\lambda ^n$ are constructed, as in
the previous example, through introducing
 vector-like $Q_{2N}$ quark doublets
$Q_i(2_{(i)}) (i=6, 7, 10, 18, 23),$
singlets
$U_i(2_{(i)}) (i=6, 10, 18)$
and $D_i(2_{(i)}) (i=4,11),$
 together with standard model
singlet Higgses
$S_i(2_{(i)}),(i= 5, 8, 11, 15, 16, 18, 23)$. As mentioned in my first
example this set of doublets which seems long and {\it ad hoc} is really
highly constrained. For this second example, the flavor group is $Q_{48}$.

Under the same kind of state labelling, the $H_u$ VEV then allows only the four
couplings:
\[ t_L<H_u>U_{6+}, \; Q_{6+}<H_u>t_R,\]
\[ Q_{6+}<H_u>U_{10+}, \; Q_{10+}<H_u>U_{6+};\]
and the $H_d$ VEV only the coupling:
\[	Q_{7+}<H_d>D_{4+}.\]
The $S_i$ VEVS may then be chosen to give certain vertices such as :
$U_{6+}^{\dag}<S_{8+-}>c_R,\\
\; U_{10+}^{\dag}<S_{8+-}>c_R$, and so on.

The general procedure is as follows: the quarks and leptons has to come from
$SU(2)_H$ singlets, doublets and triplets. There are only 21 anomaly free
schemes of
assignment with no extra chiral fermions\cite{SU2H}. I am aiming at assignments
that
can lead to up- and down-quark mass matrices with different hierarchical
textures\cite{RRR}. That leaves me with two schemes of which only the one used
here gives
interesting models.

 Picking the above
scheme, the feasiblity of using the Froggatt-Nielsen
mechanism enforces the first family quarks to be $Q_{2N}$
singlets. I then introduce appropriate heavy fermions and Higges whenever
necessary as I go on to build entries of higher order in $\lambda$,
keeping track of overall consistency.

Attempts to construct models giving texture models I, II, and III of ref.
\cite{RRR} lead to conflicts, and I therefore conclude
that those patterns of texture zeros are disfavored.

In this approach, the standard model gauge group $G =SU(3)_C \times SU(2)_L
\times
U(1)_Y$ is extended to $G \times (Q_{2N})_{global}$ which, in turn,
is assumed a subgroup of $G \times SU(2)_{H}$ where $SU(2)_H$ is
a gauged horizontal symmetry. This last point is important because
the imposition of the necessary anomaly cancellation restricts the
assignment of the quarks and leptons to $Q_{2N}$ representations as discussed
above.

The authors of [11] have analyzed all possible symmetric quark mass matrices
with the maximal (six) and next-to-maximal (five) number of texture zeros, and
concluded that only five models, denoted by the roman numerals I to V
in their work, are phenomenologically viable. By insisting on
derivation of the texture zeros from the $Q_{2N}$ dicyclic flavor
symmetry, I have reduced the number of candidates to
two.

Similar considerations can be made at the $SU(5) \times SU(5) \times Q_{2N}$
level, and this - together with the generalization to supersymmetry -
is discussed in [22] and [23].

In conclusion, the reduction in the number of free parameters in the low energy
theory attained by postulating texture zeros in the fermion mass matrices
has been shown to have a dual description in terms of a horizontal
symmetry $Q_{2N} \subset SU(2)_H$. This $SU(2)_H$ could arise in a GUT group
or directly from a superstring. My main point is that the derivation of
the values of the fermion masses and quark mixings in a putative
theory of everything may likely involve a horizontal
symmetry, probably gauged, as an important intermediate step. The simple
cases given in this talk illustrate how this can happen.\\

\bigskip
\bigskip

{\bf Acknowledgement}

\bigskip

I have been fortunate to work on different aspects of this material with T.
Kephart, O. Kong,
P. Krastev, J. Liu and D. Ng. This work was supported in part by the
U.S. Department of Energy under Grant DE-FG05-85ER-40219, Task B.\\

\bigskip

\bigskip
\bigskip

{\bf Figure Captions.}\\

\bigskip

\bigskip

Fig.1  Froggatt-Nielsen tree graphs for $M_u$. (The symmetric counterpart
$(M_u)_{23}$,
and second graphs for $(M_u)_{22}$ and $(M_u)_{33}$ are not shown).\\

\bigskip

Fig.2  Froggatt-Nielsen tree graphs for $M_d$.

\newpage

\begin{figure}[h]

\vspace*{2.0cm}

\setlength{\unitlength}{1.0cm}

\begin{picture}(15,15)

\thicklines

\put(1,13.5){\framebox(2.3,1){$(a)$: $(M_u)_{32}$}}
\put(7.08,12){\vector(1,0){1}}
\put(8.08,12){\line(1,0){0.84}}
\multiput(9,12)(0,0.3){9}{\line(0,1){0.25}}
\put(8.85,14.6){${\bf \times}$}
\put(9.08,12){\line(1,0){0.84}}
\put(10.92,12){\vector(-1,0){1}}
\multiput(11,12)(0,0.3){9}{\line(0,1){0.25}}
\put(10.85,14.6){${\bf \times}$}
\put(11.08,12){\line(1,0){0.84}}
\put(12.92,12){\vector(-1,0){1}}
\put(6.5,11.9){$t_L$}
\put(9.2,14.6){$ \left \langle {H_u} \right \rangle $}
\put(10.2,11.5){$U_{6+}$}
\put(11.2,14.6){$ \left \langle {2_8} \right \rangle $}
\put(13,11.9){$c_R$}

\put(1,8.5){\framebox(2.3,1){$(b)$: $(M_u)_{22}$}}
\put(7.08,7){\vector(1,0){1}}
\put(8.08,7){\line(1,0){0.84}}
\multiput(9,7)(0,0.3){9}{\line(0,1){0.25}}
\put(8.85,9.6){${\bf \times}$}
\put(9.08,7){\line(1,0){0.84}}
\put(10.92,7){\vector(-1,0){1}}
\multiput(11,7)(0,0.3){9}{\line(0,1){0.25}}
\put(10.85,9.6){${\bf \times}$}
\put(11.08,7){\line(1,0){0.84}}
\put(12.92,7){\vector(-1,0){1}}
\put(4.5,6.9){$c_L$}
\put(9.2,9.6){$ \left \langle {H_u} \right \rangle $}
\put(10.2,6.5){$U_{6+}$}
\put(11.2,9.6){$ \left \langle {2_8} \right \rangle $}
\put(13,6.9){$c_R$}
\put(5.08,7){\vector(1,0){1}}
\put(6.08,7){\line(1,0){0.84}}
\multiput(7,7)(0,0.3){9}{\line(0,1){0.25}}
\put(6.85,9.6){${\bf \times}$}
\put(7.2,9.6){$ \left \langle {2_8} \right \rangle $}
\put(7.5,6.5){$Q_{10+}$}

\put(1,3.5){\framebox(2.3,1){$(c)$: $(M_u)_{13}$}}
\put(7.08,2){\vector(1,0){1}}
\put(8.08,2){\line(1,0){0.84}}
\multiput(9,2)(0,0.3){9}{\line(0,1){0.25}}
\put(8.85,4.6){${\bf \times}$}
\put(9.08,2){\line(1,0){0.84}}
\put(10.92,2){\vector(-1,0){1}}
\multiput(11,2)(0,0.3){9}{\line(0,1){0.25}}
\put(10.85,4.6){${\bf \times}$}
\put(11.08,2){\line(1,0){0.84}}
\put(12.92,2){\vector(-1,0){1}}
\multiput(13,2)(0,0.3){9}{\line(0,1){0.25}}
\put(12.85,4.6){${\bf \times}$}
\put(13.08,2){\line(1,0){0.84}}
\put(14.92,2){\vector(-1,0){1}}
\put(6.5,1.9){$t_L$}
\put(9.2,4.6){$ \left \langle {H_u} \right \rangle $}
\put(10.2,1.5){$U_{6+}$}
\put(11.2,4.6){$ \left \langle {2_8} \right \rangle $}
\put(15,1.9){$u_R$}
\put(12.2,1.5){$U_{14-}$}
\put(13.2,4.6){$ \left \langle {2_{14}} \right \rangle $}

\end{picture}

\caption{Froggatt-Nielsen tree graphs for $M_u$. (The symmetric
counterpart $(M_u)_{23},$ and second graphs for $(M_u)_{22}$ and
$(M_u)_{31}$ are not shown)}

\label{Fig. 1}

\end{figure}

\newpage

\begin{figure}[h]

\vspace*{2.0cm}

\setlength{\unitlength}{1.0cm}

\begin{picture}(15,15)

\thicklines

\put(1,13.5){\framebox(2.3,1){$(a)$: $(M_d)_{22}$}}
\put(7.08,12){\vector(1,0){1}}
\put(8.08,12){\line(1,0){0.84}}
\multiput(9,12)(0,0.3){9}{\line(0,1){0.25}}
\put(8.85,14.6){${\bf \times}$}
\put(9.08,12){\line(1,0){0.84}}
\put(10.92,12){\vector(-1,0){1}}
\multiput(11,12)(0,0.3){9}{\line(0,1){0.25}}
\put(10.85,14.6){${\bf \times}$}
\put(11.08,12){\line(1,0){0.84}}
\put(12.92,12){\vector(-1,0){1}}
\put(4.5,11.9){$s_L$}
\put(9.2,14.6){$ \left \langle {H_d} \right \rangle $}
\put(10.2,11.5){$D_{4+}$}
\put(11.2,14.6){$ \left \langle {2_5} \right \rangle $}
\put(13,11.9){$s_R$}
\put(5.08,12){\vector(1,0){1}}
\put(6.08,12){\line(1,0){0.84}}
\multiput(7,12)(0,0.3){9}{\line(0,1){0.25}}
\put(6.85,14.6){${\bf \times}$}
\put(7.2,14.6){$ \left \langle {2_5} \right \rangle $}
\put(7.5,11.5){$Q_{7+}$}

\put(1,8.5){\framebox(2.3,1){$(b)$: $(M_d)_{21}$}}
\put(7.08,7){\vector(1,0){1}}
\put(8.08,7){\line(1,0){0.84}}
\multiput(9,7)(0,0.3){9}{\line(0,1){0.25}}
\put(8.85,9.6){${\bf \times}$}
\put(9.08,7){\line(1,0){0.84}}
\put(10.92,7){\vector(-1,0){1}}
\multiput(11,7)(0,0.3){9}{\line(0,1){0.25}}
\put(10.85,9.6){${\bf \times}$}
\put(11.08,7){\line(1,0){0.84}}
\put(12.92,7){\vector(-1,0){1}}
\put(4.5,6.9){$s_L$}
\put(9.2,9.6){$ \left \langle {H_d} \right \rangle $}
\put(10.2,6.5){$D_{4+}$}
\put(11.2,9.6){$ \left \langle {2_{13}} \right \rangle $}

\put(5.08,7){\vector(1,0){1}}
\put(6.08,7){\line(1,0){0.84}}
\multiput(7,7)(0,0.3){9}{\line(0,1){0.25}}
\put(6.85,9.6){${\bf \times}$}
\put(7.2,9.6){$ \left \langle {2_5} \right \rangle $}
\put(7.5,6.5){$Q_{7+}$}
\multiput(13,7)(0,0.3){9}{\line(0,1){0.25}}
\put(12.85,9.6){${\bf \times}$}
\put(13.08,7){\line(1,0){0.84}}
\put(14.92,7){\vector(-1,0){1}}
\put(15,6.9){$d_R$}
\put(12.2,6.5){$D_{17-}$}
\put(13.2,9.6){$ \left \langle {2_{17}} \right \rangle $}

\put(1,3.5){\framebox(2.3,1){$(c)$: $(M_d)_{12}$}}
\put(7.08,2){\vector(1,0){1}}
\put(8.08,2){\line(1,0){0.84}}
\multiput(9,2)(0,0.3){9}{\line(0,1){0.25}}
\put(8.85,4.6){${\bf \times}$}
\put(9.08,2){\line(1,0){0.84}}
\put(10.92,2){\vector(-1,0){1}}
\multiput(11,2)(0,0.3){9}{\line(0,1){0.25}}
\put(10.85,4.6){${\bf \times}$}
\put(11.08,2){\line(1,0){0.84}}
\put(12.92,2){\vector(-1,0){1}}
\put(2.5,1.9){$d_L$}
\put(9.2,4.6){$ \left \langle {H_d} \right \rangle $}
\put(10.2,1.5){$D_{4+}$}
\put(11.2,4.6){$ \left \langle {2_5} \right \rangle $}
\put(13,1.9){$s_R$}
\put(5.08,2){\vector(1,0){1}}
\put(6.08,2){\line(1,0){0.84}}
\multiput(7,2)(0,0.3){9}{\line(0,1){0.25}}
\put(6.85,4.6){${\bf \times}$}
\put(3.08,2){\vector(1,0){1}}
\put(4.08,2){\line(1,0){0.84}}
\multiput(5,2)(0,0.3){9}{\line(0,1){0.25}}
\put(4.85,4.6){${\bf \times}$}
\put(7.5,1.5){$Q_{7+}$}
\put(5.5,1.5){$Q_{13-}$}
\put(7.2,4.6){$ \left \langle {2_{20}} \right \rangle $}
\put(5.2,4.6){$ \left \langle {2_{13}} \right \rangle $}

\end{picture}

\caption{Froggatt-Nielsen tree graphs for $M_d$. }

\label{Fig. 2}

\end{figure}


\bigskip
\bigskip
\newpage


\bigskip


{\bf References}

\bigskip

\begin{thebibliography}{999}

\bibitem{PHF}
The original reference is P. H. Frampton, Phys. Rev. Lett. {\bf
69}, 2889 (1992);
The treatment of neutrinos is in P. H. Frampton, P. Krastev and J. T. Liu,
Mod. Phys. Lett. {\bf 9A}, 761 (1994);
The phenomenology has appeared in P. H. Frampton, J. T. Liu, D. Ng and
B. C. Rasco, Mod. Phys. Lett. {\bf 9A}, 1975 (1994);
The suggestion of a dilepton resonance in $e^-e^-$ was first mentioned
in P. H. Frampton and B. H. Lee, Phys. Rev. Lett. {\bf 64}, 619 (1990).
\bibitem{books}
Useful sources of information on the finite groups include:\\
D. E. Littlewood, {\bf The Theory of Group Characters and Matrix Repesentations
of Groups}, Oxford (1940).\\
M. Hamermesh, {\bf Group Theory and Its Applications to Physical Problems},
Addison-Wesley (1962).\\
J. S. Lomont, {\bf Applications of Finite Groups}, Academic Press (1959),
reprinted by Dover (1993).\\
A. D Thomas and G. V. Wood, {\bf Group Tables}, Shiva Publishing (1980).
\bibitem{guts}
Another route to discrete family
symmetry is through the previous $SU(N)$ family symmetry models\cite{SUN}
. Here one can consider the breaking pattern
\begin{equation}
SU(N)\rightarrow SU(5)\times SU(5-N)\times U(1)\rightarrow SU(5)\times G
\end{equation}
where G is a discrete group. Such models typically have more than just
three complete families until $G$ is broken.
\bibitem{SUN}
H. Georgi, Nucl. Phys. {\bf B156}, 126 (1979);\\
P. H. Frampton, Phys. Lett. {\bf 88B}, 299 (1979).
\bibitem{Pak}
S. Pakvasa and H. Sugawara, Phys. Lett. {\bf 73B}, 61 (1978); ibid {\bf 82B},
105 (1979);\\
T. Brown, N. Deshpande, S. Pakvasa and H. Sugawara, {\it ibid} {\bf 141B}, 95
(1984);\\
T. Brown, S. Pakvasa, H. Sugawara and Y. Yamanaka, Phys. Rev. {\bf D30}, 255
(1984).
\bibitem{global}
S. Giddings and A. Strominger, Nucl. Phys. {\bf B307}, 854 (1988);\\
S. Coleman, {\it ibid} {\bf B310}, 643 (1988);\\
G. Gilbert, {\it ibid} {\bf B328}. 159 (1989);\\
R. Holman, S. D. H. Hsu, T. W. Kephart, E. W. Kolb, R. Watkins and L. M.
Widrow,
Phys. Lett. {\bf B282}, 132 (1992);\\
M. Kamionkowski and J. March-Russell, {\it ibid} {\bf 282}, 137 (1992);\\
S. M. Barr and D. Seckel, Phys. Rev. {\bf D46}, 539 (1992).
\bibitem{flat}
"Invariants of Flat Bundles" by J. Cheeger, Proceedings of the
International Congress of Mathematicians (Vancouver, 1974)
"Multiplication of Differential Characters" by J. Cheeger, Instituto
Nazionale di Alta Matematica, Symposia Mathematica, Volume XI (1973)
"Differential Characters and Geometric Invariants" by J. Cheeger and
\bibitem{GUTS1}
S. Dimopoulos, L. J. Hall and S. Raby, Phys. Rev. Lett.
{\bf 68}, 1984 (1992); Phys. Rev. {\bf D45}, 4195 (1992); {\it ibid.} {\bf
D46},
R4793 (1992);\\
G. Anderson, S. Raby, S. Dimopoulos, and L. J. Hall, {\it ibid} {\bf D47}, 3702
(1993)
\bibitem{GUTS2}
G.F. Giudice, Mod. Phys. Lett. {\bf A7}, 2429 (1992).
\bibitem{GUTS3}
G.K. Leontaris and N.D. Tracas, Phys. Lett. {\bf B303}, 50
(1993);\\
 G.K. Leontaris and J.D. Vergados, {\it ibid}. {\bf B305}, 242 (1993).
\bibitem{RRR}
P. Ramond, R.G. Roberts and G.G. Ross, Nucl. Phys. {\bf 406}, 19 (1993).
\bibitem{U1P}
E. Papageorgiu, Z. Phys. {\bf C64}, 509 (1994).
\bibitem{U1}
L. Ibanez and G. G. Ross, Phys. Lett. {\bf B332}, 100 (1994).
\bibitem{finite1}
S. Pakvasa and H. Sugawara, Phys. Lett. {\bf 73B}, 61
(1978); {\it ibid}. {\bf 82B}, 105 (1979).
\bibitem{finite2}
D.B. Kaplan and M. Schmaltz, Phys. Rev. {\bf D49}, 3741 (1994).
\bibitem{finite3}
P. H. Frampton and T. W. Kephart, Phys. Rev. {\bf D51}, R1 (1995), and
Int. J. Mod. Phys. {\bf A} (1995, in press).
\bibitem{FN}
C. D. Froggatt and H. B. Nielsen, Nucl. Phys. {\bf B147}, 277 (1979).
\bibitem{nonsym}
M. Leurer, Y. Nir and N. Seiberg, Nucl. Phys.{\bf B398}, 319 (1993); {\it
ibid}. {\bf B420}, 468 (1994);
E. Dudas, S. Pokorski and C.A. Savoy, {\it Yukawa Matrices from a Spontaneously
Broken Abelian Symmetry}. preprint SPhT Saclay T95/027, MPI-PTh 95-33
\bibitem{EFT}
H. Georgi, Nucl. Phys. {\bf B} (Proc. Suppl.) {\bf29B,C}, 1 (1992).
\bibitem{SU2H}
D.S. Shaw and R.R. Volkas, Phys. Rev. {\bf D47}, 241 (1993).
\bibitem{FK1} P.H. Frampton and O.C.W. Kong, Phys. Rev. Lett. {\bf 75}
(1995, in press)
\bibitem{FK2}
P. H. Frampton and O. C. W. Kong, UNC-Chapel Hill Report No.
IFP-715-UNC (1995, in preparation).
\bibitem{Kong}
O.C.W.Kong, these proceedings.

\end{thebibliography}
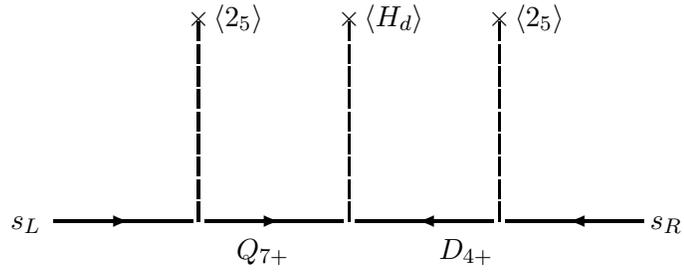
\end{document}